\documentclass[12pt]{article}
\usepackage{epsfig}

\usepackage{amsmath}
\usepackage{amssymb}
\usepackage{euscript}

\begin{document}

\begin{titlepage}

\begin{center}
\vspace{25mm}
{\LARGE\bf Exploring confinement} \\
\end{center}

\vspace{25mm}


\begin{center}
{\large\bf  Mieczyslaw Witold Krasny } \\

\vspace{5mm}

LPNHE, Universities Paris VI et VII, IN2P3-CNRS, Paris \\

\end{center}

\vspace{60mm}
\begin{abstract}
This note is an extended version of the contribution to the CERN Council Open Symposium on 
European Strategy for Particle Physics. It discusses an experimental  programme to explore the QCD confinement phenomena at CERN
with a  new electron-proton  and electron-nucleus collider using  the existing SPS beams (optionally also the future  SPL and PS  proton and ions beams) and the polarised electron beam in the range of 5 to 20 GeV  from a newly built Energy Recovery Linac. 

\end{abstract}


\end{titlepage}

\newpage

\section{Introduction}

While the elementary,  point-like  particles, carrying a unit electric charge 
(expressed in terms  of the electron charge) can propagate freely in vacuum, those   
carrying fractional charges: quarks,  cannot.  They are confined,  in the 
present day Universe, within fermi-size objects: nucleons.
        
The strong force which confines quarks, does not only  
allow to store their relict, early Universe,  kinetic energies  
in the form of the mass of the nucleons but, in addition,  
provides the mechanism for a gradual,  life-sustaining  release of a  small fraction of the stored energy.  
This  mechanism is  the process of formation of atomic nuclei  taking place in the core of the stars.

Given the importance of the strong interactions it is hard to admit  that,  precisely 100 years after 
the discovery of atomic nucleus,  the role of the quark degrees of freedom  in the strong interactions phenomena involving nucleons and nuclei,  remains still to be investigated experimentally and understood. 
 
We hardly understand the mechanism which confines quarks in nucleons,  nor the role of the quark degrees of freedom in forming atomic nuclei. We do not know if the gauge carriers of strong interaction forces: gluons, can propagate in atomic nucleus, or whether they are confined within nucleons.  Moreover,  we do not know if they can propagate within the nucleon volumes,  or whether they are confined within the instanton or constituent quark distance scales. 

We have still not fully understood  the  quark/gluon orbital momentum in hadrons and  have no idea  how the movements of quarks and gluons are correlated within nucleon/nucleus. We do not know how  ``rigid" the nucleon and nuclear matter is,  i.e.,   how easy/difficult is to pull/push or rotate  the nucleons, placed  in the vacuum and in the nuclear matter,  by puling/pushing or spinning of one  of its quarks or gluons.

It is fair to say that  over the last 40 years the domain of our ignorance was partially reduced. 
A significant  experimental and theoretical progress
has been made in understanding the strong force at the distances sizeably smaller that 
the quark confinement distances,  where it strength weakens. 
Quantum Chromodynamics (QCD) has been established to be an
underlying theory of interactions of quarks and gluons.  
Its quantitative predictions based upon the perturbative calculation techniques 
passed successfully the experimental data scrutiny. 
More advanced computational techniques (lattice calculations) have been subsequently 
developed to extend the predictive power of QCD to static observables such as the masses 
of baryons and mesons.   

Despite of all the above successes, the basic observables describing  the quark and gluon dynamics at the confinement scale are beyond the jurisdiction of the  present day calculation methods of QCD. For example: the partonic momentum distributions in the nucleus and in the nuclei at the fixed resolution  scale  cannot be calculated/predicted, neither perturbatively nor using 
the  lattice techniques. The fragmentation functions of quarks into hadrons both in the vacuum and in the hadronic matter are also beyond the reach of the available computational methods of the QCD. 

The present day QCD status resembles that of the Quantum Electrodynamics (QED)  in the first halt of the 20th century:  many 
intriguing  observed phenomena could not have been derived from the QED Lagrangian alone. 
What is striking is  a remarkable  difference in the respective  experimental programs:
the lack of adequate QED theoretical tools did not stop,   but rather accelerated or strengthen,  the curiosity driven experimental investigation of  the media composed of atoms in all its aspects including those which escaped the jurisdiction of the rigorous calculation methods. Such a curiosity  driven research led to milestone  discoveries which could have never been predicted by the theory.   Superconductivity and superfluidity, both lying foundations for  the  construction of the present day particle accelerators,  are the notable examples here. 

On the contrary, the QCD confinement-focused experimental programme has hardly been addressed.  Some of its aspects have been studied  at the Jefferson Lab, BNL, CERN and DESY.   However, to a large extent,  the confinement phenomena have been considered more as a burden than as the research target. As a consequence the quest for understanding was often replaced by an efficient absorbing  the lack of knowledge in terms of  phenomenological models or parametrisations,  with  plethora of ad-hoc parameters having  no link  the underlying theory of strong interactions.

\section{The past and the present context}

\subsection{DESY }

A confinement-focused experimental programme has been proposed in 1996 \cite{DESY_workshop}, \cite{memorandum}  as one of  the three possible extensions of the HERA programme. The necessary upgrades of the DESY accelerator infrastructure, in particular the construction of the new proton and ion injectors  jointly by GSI and DESY were  discussed at the Seeheim workshop \cite{Seeheim}. 

Unfortunately, this programme was proposed at the time when DESY was aiming to build TESLA, and since only one of these two above options could be pursued, the confinement project was abandoned, leaving a place for the TESLA project. The  "high" luminosity programme, the least interfering with TESLA,  was chosen as an extension of the HERA programme at DESY.  

When TESLA project was finally abandoned,  GSI had already embarked on the development of the FAIR project. The opportunity for DESY to become the leading world laboratory for the studies of confinement phenomena in terms of the quark and gluon degrees of freedom had been lost.

\subsection{BNL} 

A  reincarnation attempt of the confinement-focused experimental programme,  tailored to the BNL accelerator infrastructure,  was made in  the years 1999-2001. The first ideas for the BNL based experimental facility for  confinement studies  were presented at the Moriond meeting in March 1999 \cite{Moriond1999}. It was followed by the first design of the eRHIC collider \cite{first_physics _white_paper} and by the first design of the full acceptance detector specialised in  the confinement research programme \cite{krasny_yale}, . 

At the 2001 Snowmass workshop, devoted to the ``Future of the High Energy Physics",  the  role of intermediate energy electron-proton  and electron-ion colliders for the confinement studies was discussed. The optimisation of the collider and detector parameters to address the confinement programme was  summarised in   \cite{krasny_physics} and \cite{krasny_detector}. 

The initial momentum of the BNL-based confinement project has been significantly reduced in 2002  by the decision of the NSAC long range planning committee  to  put the Rare Isotope Accelerator (RIA), presently called the Facility for Rare Isotope Beams (FRIB),   as the first priority  project on the  list of the Nuclear Physics large infrastructure projects in the USA. Nevertheless, the eRHIC collider  design has been, since then,  refined \cite{eRHIC}.  In addition the  ELIC project at TJNAF have been proposed \cite{ELIC}. These two accelerator projects are presently competing to be endorsed by the NSAC as the highest priority project  after  a completion of the JLAB 12 GeV upgrade and after a  completion of FRIB.

\subsection{CERN} 

Recently,  a project of colliding the LHC proton and ion beams with the beam of electrons was resurrected at CERN and the design report of the LHeC collider has been presented \cite{LHeC}. 

This initiative, even if driven by different research  targets,  shares the common interest with the programme discussed in this note in creating at CERN the high intensity polarised electron beams. 

\section{The experimental programme}

The research target of the confinement programme  is to  investigate experimentally quark and gluon dynamics at the confinement distance scale by studying the response of the  variable colour-charge configuration QCD media  -- vacuum, nucleons and nuclei --  to experimentally well resolved and theoretically well controlled electromagnetic perturbations. As in  the previous high energy lepton scattering experiments at SLAC, CERN and HERA the initial perturbation of quark and gluon degrees of freedom  is  controlled experimentally by a high resolution measurement of the outgoing lepton momentum and theoretically, in the restricted domain of the four momentum transfer to the coloured medium, by the perturbative QED. However, the  lepton scattering process is  used as the surgery tool rather than the research target. 

The confinement programme extends the the past QCD research in the following aspects:

\begin{enumerate}

\item 
The adjustable focal  length of the lepton probe would allow, for the first time in a single experiment,  to  cover the full dynamical range required for studies of confinement phenomena: the distances of 0.01 - 10 fermi in the direction transverse to the colliding particle axis, and the distances of 0.01 -100 fermi in the longitudinal direction. The low ,"calibration",  limit assures applicability of the leading twist QCD perturbative techniques to control   the response of the strongly interacting matter to the EM perturbation. The upper limit assures that all the relevant strong interaction length scales are covered (constituent quark size, nucleon size, nuclear size).  

\item
On top of the proton  beams,  a  broad range on ion beams are foreseen --  both the isoscalar beams, such as deuterium, helium, oxygen and calcium ions and the large atomic number beams, such as the lead ion beams.

\item 
The nuclei would  no longer  play  only the role of passive targets (as in the previous fixed target DIS experiments) but also the role of femto-detectors to study the space-time dependent aspects of the strong interactions  at the confinement distance scale.

\item 
The confinement programme would make a full profit from the tagged momentum photon beams both for the studies of photon initiated nucleus disintegration processes, but also for the creation of strongly interacting matter in photon-photon collisions.

\item 
It could provide the necessary input  measurements for the LHC experimental programme   to either significantly improve  
the LHC measurement precision (e.g. to measure the sea/valence structure of the proton for high precision measurements of the electroweak parameters at the LHC) or to avoid the interpretation ambiguities of the LHC results ( e.g. resolving the initial and the final state interactions in the  hard AA collisions). 
 
\end{enumerate}

\section{The research facility}

The  facility to conduct the confinement programme is a  specialised  lepton-proton  and lepton-ion  collider. Two  complementary detectors are needed in these studies: (1) a large $\beta ^*$ beam crossing  detector capable to detect all the particles produced in the collisions and to  resolve the nuclear fermi-motion scale  momenta of all the nucleons and nuclear fragments,   and (2) a small  
$\beta ^*$ detector optimised to achieve the highest statistical precision of selected observables, in particular for the studies of rigidity of hadronic matter, for investigation of the spin and orbital momentum of nucleon constituents and for the studies of quark and gluon momentum and angular momentum correlations.

\section{The collider parameters}

\subsection{The centre-of-mass energies}

This programme requires a broad range of the collision centre-of-mass energies. 
The optimal energy range  is specified by the following boundary conditions: 10 GeV $\leq $ $\sqrt{s}$ $\leq$  200 GeV  \cite{krasny_physics}. 
The low limit allows for a significant overlap with the TJNF high luminosity 
fixed-target measurements. The upper limit assures an overlap with the HERA measurements. 
The proposed range covers the requisite EM-probe resolution range of the longitudinal and transverse distances 
and allows to separate the processes of absorption  of  the transversely and longitudinally polarised virtual photons in  hadronic mater.

\subsection{Luminosity} 

The luminosity range: 
$10^{30}$ cm$^{-2}$s$^{-1}$ $\leq \cal{L} \leq$ $10^{33} $ cm$^{-2}$s$^{-1}$ is optimal  
for the experimental programme discussed in this note. While the low luminosity runs are  sufficient for precise studies of the nucleus disintegration processes, the highest ones are  necessary  to measure the polarisation asymmetries  and multidimensional quark fragmentation functions in vacuum and in nuclear media with a high statistical precision. A possible running scheme is to 
use the low emittance beams and to tune the value of the $\beta ^*$ for the optimal  trade-off of the luminosity and measurement precision performance.  

\subsection{Flavour, charge, and polarisation of the lepton beam} 

The main  reason to chose the electron beam  instead of the muon beam  is to reach  the luminosity targets specified above.  
The price to pay is a more sophisticated  insertion of the beam and the final focussing in the Interaction Point (IP) 
to handle the photon  radiation in the beam collisions zone
(this can be partially circumvented by choosing highly asymmetric energies  for the nucleon and for the electron beam, as discussed later in this section).  

Electron and positron beams are equivalent for the proposed programme. Polarisation of the 
electron/positron  beam is important, however not critical. 

\subsection{The choice of ions}

The highest atomic number A  ions and the light isoscalar ions are of comparable importance. 
While for the e-Pb  collisions   the main emphasis would  be on maximising the medium effects, 
the isoscalar beams such as He, O or Ca would  be beneficial   for the  high-precision relative measurements -- 
in particular,  if the maximum momentum spread of the stored ions of 0.25 \% of the nominal value can be tolerated 
and  the  He, O and Ca ion bunch trains could collide simultaneously. 

A special emphasis would be on the runs with D beam.

\subsection{The ratio of the proton(ion) and electron beam energies}

There are several reasons to choose a highly  asymmetric, $E_{nucl} \gg  E_e$,  collision scheme: 
\begin{itemize} 
\item
better angular separation of the scattered quark associated particles and nuclear fragments,
\item
better resolution power of the electromagnetic probe,
\item 
easier recognition of diffractive events,
\item
decoupling of the electron and the ion  IP beam optics, 
\item
detection, identification and measurement of all  products of nuclear fragmentation.
\end{itemize} 

The upper limit of the maximal energy of the nucleon is specified by the requirement 
of detecting of all the nuclear fragments: evaporated nucleons, wounded nucleons,  and nuclear de-excitation gammas. 
Studies presented at the 2001 Snowmass workshop showed that above the energy of $\approx$200 GeV/nucleon 
the design of the IR beam insertion and focusing optics was in conflict with 
the requirement to  identify nuclear fragments and to measure their  momenta with the requisite precision. 
The lowest nucleon energy for which the atomic number of the detected nuclear fragment can be unambiguously 
identified using the measurement of the deposited energy in the calorimeter is $\approx$4 GeV. 

For the electron beam momentum smaller than $\approx$5 GeV the identification issues of the scattered electron 
in highly inelastic events become critical. This defines the lower limit of the electron beam energy. 
The upper limit  is constrained by the radiation power and the critical energy value in the 
IP zone.  Note that a small-bend electron insertion is incompatible with a full acceptance capacity of the detector. 
All the above constraints provide a rather sharp upper limit of the maximum electron beam energy of  $\approx$20 GeV  \cite{krasny_detector}.

\section{The confinement project at CERN}

The eRHIC project \cite{first_physics _white_paper}  which adds to  the existing BNL RHIC ring  the Energy-Recovery-Linac (ERL) fulfils most of the requirements discussed in the previous section. 
However,  the eRHIC project was proposed already more than 10 years ago and as the time passes  it becomes more and more unlikely that it  will ever be realised at the BNL site. In addition,  its target physics programme drifted away from the initial confinement project towards a mere continuation of the HERA programme at the 
higher luminosity machine.  The status of the lower energy 
TJNAF  ELIC project \cite{ELIC} is similar to that of the eRHIC project.
If ever constructed it will  come late, certainly after the TJNAF  12 GeV upgrade. 

The proposal presented here is to implement  the confinement research programme at CERN in synergy with the planned upgrade of the LHC injectors: SPL and PS and with the ongoing LHC physics programme.
The basic  idea is to construct,  at the CERN site,  an Energy Recovery Linac (ERL) to accelerate 
polarised electrons to the top energies in the range of 5-20 GeV and to collide the ERL beam with the proton 
and ion beams stored in  the SPS (in the equivalent proton energy range of 170 - 400 GeV). 
In addition,  delivering the future SPL proton beam and the PS proton and ion beams at their  top energies to the electron-proton and electron-ion  collision interaction point(s) IPs would  be highly beneficial for the proposed physics programme. 
The ERL design could e.g. follow that of the eRHIC collider based on the 4 pass energy recuperation scheme an providing electron  beams of the energies of 5, 10, 15 and 20 GeV.

The main advantages of the ERL based electron-proton and electron-ion collider scheme are:
\begin{itemize} 
\item
large luminosity (the electron bunch interacts only once), 
\item 
an easy variation of the electron bunch frequency adjusted to the proton/ion bunch frequency at variable proton/ion energies (note, that ions and protons in the large fraction of the proposed energy range are not fully relativistic),
\item
very long ``free" straight section in the vicinity of the IP allowing to design  a 4$\pi$ detector capable to measure 
the products of the nucleus disintegration (the femto-detector signals),  
\item 
high (80 \%) electron beam polarisation at each of the electron beam energies,
\item 
a low emittance electron beam
\end{itemize} 
   
The three LHC injectors: SPL, PS and SPS,  together with  the ERL could provide the full range  of the optimal electron and hadron beam energies for the confinement research programme. 

To implement such a programme at CERN an  $R\&D$ at the present technology frontier on:
ERLs, polarised electron guns, and advanced cooling techniques of the ion beams would be required. 
The $R\&D$ in this domain could be useful non only for the confinement project but also for the future high energy project at CERN: the CLIC project. 
As far as the accelerator technological challenges are concerned,  the confinement project is equivalent to the 
LHeC project.  But the analogy ends here. In all the other aspects the confinement project targets differ from those of the LHeC programme.

\section{The LHeC and the confinement programme} 
 
\subsection{The physics} 

The confinement programme cannot be conducted at the LHeC. 
The physics goals of the LHeC and of the confinement proposals are distinct.

The LHeC can be considered as the continuation of the HERA scientific programme 
at the high energy frontier and its merits should be judged in comparison with the other 
high energy frontier projects. The comparison of the LHeC with LHC is particularly straightforward. 
It can be extrapolated from the relative merits of the HERA and the Tevatron projects because:  (1) HERA and Tevatron  were operating at similar proton beam energy (as in the case of the LHeC and LHC), (2)  the ratio of the HERA electron beam energy to the proton beam energy is similar 
as in the LHeC case, and (3) the ratio of collected luminosities at HERA and 
at the Tevatron are similar to the  ratio of the design values of luminosities collected by the LHeC and the LHC. 

It is rather obvious that HERA hardly improved our knowledge of the electroweak (EW) sector of the Standard Model. In the QCD sector,  it provided a very  important initial input to the LHC programme by measuring some combination of the  quark parton distribution functions (PDFs)
and by deriving the gluon PDF from the scaling violation of the DIS structure functions. 
However, the precision achieved  at HERA  turned   out to be inferior with respect  to the one required for 
improving the present precision EW measurements  at the LHC \cite{Krasny_MW} (mostly due to large statistical errors of the charged current and heavy quark cross section  measurements). The requisite  precision target can certainly be reached at the LHeC. Unfortunately such an improvement would  come  out of phase with the LHC programme - it is unlikely that the LHC results will be reanalysed.

The confinement programme discussed here would be  focussed on asking new questions and exploring new territories rather than on continuing the measurements of canonical  observables in the new energy regime.  It would be complementary to the QCD research programs at TJNAF and at the future FAIR facility at Darmstadt. It could provide also a necessary  input data for: (1)  increasing the precision of the LHC EW measurements (precision input information  for understanding of the W and Z boson polarisation at the LHC) and   (2) experimental  resolving of the final and initial state state effects in the hard  AA collisions   --   both in phase with the ongoing LHC experimental programme.

\subsection{The cost} 

The cost of the confinement project represents a small fraction of the LHeC project due to two important factors:
(1) the requisite maximal energy of the electron beam is smaller by factor of 3-10 than in the  LHeC case; 
(2) the electron ring could be placed in the SPS tunnel  (as foreseen already in the year 1976 in the 
design of  the CHEEP project \cite{CHEEP}). This would significantly  reduce  the civil engineering work on the transfer lines. 

\subsection{Synergies} 

The proposed project optimises the duty cycle of the LHC injectors. As soon as  the LHC 
is filled the proton/ion bunches, the  SPS could be used for collisions with the electron beam. The interference of the confinement programme with the 
the LHC pp and AA collision programme would thus be minimal.  The construction of the ERL could  be,  to a large extent,  decoupled from  the 
ongoing LHC operation. The CERN confinement project could attract  the eRHIC and ELIC communities. CERN could thus become not only the leading world laboratory for the electroweak interaction studies  at the high energy frontier but also in the domain of strong interactions at the exploratory frontier.

\section{Conclusions}

New collider projects addressing the high energy frontier of particle physics have certainly the highest potential  in discovering new phenomena. However they require costly investments. Given the present financial crisis in  Europe,  it may be worthwhile  to consider, at present,  a relatively ``low" cost accelerator project for CERN. Such a project could be realised in parallel to the ongoing LHC experimental programme, while waiting for a clear vision for the most optimal  high energy frontier project which  can be establish  only following the completion of the high luminosity phase of the LHC scientific programme. 

The confinement project could play such a role. It could  be designed  in synergy with the upgrade 
of the LHC injectors, executed in parallel with the LHC experimental programme,   and could provide better understanding of the confinement phenomena. Moreover,  the investment in the ERL technology inherent  to this project, may be crucial in developing the next high energy frontier project at CERN.

\end{document}